\documentclass[A4]{optica-article} %%% modified margins -> \geometry{nohead,left=1.in,right=1.in,top=1.in,bottom=1.in,headheight=15pt,headsep=\dimexpr1.3in-48pt-15pt\relax} %% uses geometry.sty

\journal{opticajournal} % for journals or Optica Open

\articletype{Research Article}

\usepackage{xspace} % intelligent spacing after user-defined commands etc

\newcommand{\ndyso}[0]{{Nd$^{3+}$:Y$_2$SiO$_5$}\xspace}

\newcommand{\yso}[0]{Y$_2$SiO$_5$\xspace}
\newcommand{\er}{Er$^{3+}$\xspace}
\newcommand{\nd}{Nd$^{3+}$\xspace}

\newcommand{\bax}{{\bf b}}
\newcommand{\Du}{{\bf D$_1$}}
\newcommand{\Dd}{{\bf D$_2$}}
\begin{document}
\title{Refractive indices of \yso in the near-infrared}
\date{\today }
\author{Shijun Zhang\authormark{1}, Jérôme Debray\authormark{1}, Benoît Boulanger\authormark{1}, Pierre Lemonde\authormark{1}, Thierry Chanelière \authormark{1,*}}

\address{\authormark{1}Univ. Grenoble Alpes, CNRS, Grenoble INP, Institut N\'eel, 38000 Grenoble, France}

\email{\authormark{*}thierry.chaneliere@neel.cnrs.fr} 

%\date{\today}
%\maketitle

\begin{abstract*}
%We measure the principal values of the refractive index of \yso in the telecom range. We use the crystal itself as a Fabry-Perot etalon. The spectral fringes spacing allow to deduce the indices in different directions. This method is at the state of the art in terms of precision reaching ?? $xx.10^{-4}$.
%Our objective is to reconcile measurements in the visible and infrared by performing an independent measurement.

\yso is a reference birefringent material for optical quantum technologies. The refractive index is primarily known in the visible spectrum, whereas this crystal is also used in the near-infrared. We begin by analysing historical measurements, as well as the modelling proposed at the time. The absence of refractive index tabulated values in the near- and mid-infrared ranges motivated us to carry out independent measurements. Using interferometric techniques in the telecom wavelength range, we demonstrate the ability to determine the principal refractive indices and propose a model spanning a broad spectral range based on Sellmeier equations for which we explicitly give the coefficients and discuss the achievable accuracy.
Conversely, as an illustration, white-light interference enables us to precisely determine the thickness of a \yso thin film on the order of ten micrometers.

\end{abstract*}

\section{Introduction}
Rare-earth doped \yso has emerged as a promising materials for quantum photonics. The long coherence times observed, which are consistent with the low density of nuclear spins in the matrix, have prompted numerous spectroscopic studies and, more recently, practical demonstrations for optical quantum memories \cite{Zhong2015, Rani2017, Nicolle:21, PhysRevLett.127.210502, Sanchez_Mejia_2026} 
 and optical-microwave transduction \cite{Gonzalvo, nicolas2023coherent}. Theses optical studies with different dopants covering the visible (Eu$^{3+}$ and Pr$^{3+}$), the near-infrared (telecom) range (Er$^{3+}$). These are generally supplemented by studies of nuclear or electronic spin transitions, which provide particularly detailed knowledge of different rare earth in \yso.
 
Surprisingly, the principal values of the refractive index in the dielectric frame (called {\bf b, D$_1$, D$_2$} where {\bf b} is the monoclinic crystal axis, and {\bf D$_1$, D$_2$} lie in the crystallographic ({\bf a, c}) plane with the orientation defined by \cite{li1992spectroscopic}) are only approximately known because there is only a single measurement performed in the 430-650\,nm range  with a heavily doped \ndyso sample \cite{sellmeier} (10\% \nd concentration). Given the interest in ytterbium and erbium at 979\,nm and 1536\,nm respectively, it seems important to know the refractive indices in the infrared. We will first discuss, in section \ref{Extrapolating}, the choice of coefficients in the historical model of Beach {\it et al.} and its extrapolation to the infrared.

%especially since extrapolating Beach {\it et al.} measurements is not scientifically sound. We shall therefore begin by discussing  in section \ref{Extrapolating} the limitations of the historical model and the alternative that we consider most appropriate.

%This imprecision hampers the development of photonics devices for which an accurate knowledge of the refractive indices is necessary, for example to design single mode waveguides and more generally to properly evaluate the birefringence induced in this biaxial crystal.

%For our measurements, we focus on the telecom range for the two following reasons. Firstly because the \eryso active transition falls in the C-band and is actively investigated for deployment of quantum technologies. In this context, very weak \er concentration ($<$100ppm) are targeted to avoid spin-spin interaction \cite{dajczgewand2014optical} as opposed to the historical measurement of Beach {\it et al.} with a 10\% \nd doped sample \cite{sellmeier}. Secondarily because Beach {\it et al.} worked in the visible so the measurements in the form of Sellmeier coefficients cannot be reasonably extrapolated in the short-wave infrared.

In the telecom range, we employ a spectral interferometric technique and use the crystal itself as a Fabry-Perot etalon. As coherent source are now widely available, interferometric methods, first in the spatial domain, have been rapidly identified as an precise alternative to the historical techniques based on refractometry \cite{Singh_2002, Lee:16}. Indeed, spatial fringes produced by the sample under study can be analysed to evaluate the index of refraction. Measurements in the spectral domain have also appeared early as extremely precise, with the advantage that white light interferometry (using broadband source as opposed to lasers) allows to acquire immediately the dispersion curve \cite{SAINZ1994381, Galli:03, AROSA2022110225}. In that case, a standard interferometer (Michelson or Mach-Zehnder for example) is assembled and used as a measurement test-bed in which different samples can be inserted.

Our aim is not to perform high-precision refractometry of \yso, even if the measurement techniques have been refined and specifically improved for optical crystals. The most commonly used method is the minimum deflection angle, whose accuracy meets the requirements of nonlinear optics for calculating phase-matching wavelength ranges, up to a $10^{-6}$ precision. It requires high-quality, well-polished and large-size crystal samples (typically >10 mm in size) to achieve reliable measurements. Reference \cite[and references therein]{SHI2023100017} is a recent review and a good entry point for the literature. The technical challenges are compounded in the case of birefringent materials, for which several orientations must be investigated using different samples of equivalent qualities. In any case, the monoclinic nature of \yso imposes limitations on the precision of the index measurements.
In the section \ref{Extrapolating}, we will also discuss the accuracy of the measurements and the sources of systematic error.

Here, we directly use the millimeter-size crystal as a Fabry-Perot interferometer to complete in the infrared the historical index measurements. The relatively high index of \yso ($\sim$ 1.8 \cite{sellmeier, PhysRevB.67.245108}) is sufficient to observe well-contrasted fringes in the transmitted spectrum because of the facets reflection (Fresnel reflection). This technique doesn't require any external interferometic setup but the parallelism of the opposite faces should be ensured during the polishing process. This approach has been used naturally in the mid-infrared (large wavelength) \cite{TAN1998158, 10.1117/12.439193} or thin film samples  \cite{Brindza:14, Poelman_2003} for which a moderate spectrometer resolution is sufficient. The method is actually well-adapted to the telecom range because commercially available high resolution spectrometers ($<$0.01nm) allow to observe fringes though millimetres of propagation. In turn, millimeter-size crystals can be shape as cuboids to probe three directions with the same sample as opposed to three independent thin samples (with the risk of relative disorientation of the crystalline axis between samples). In the C-band, an erbium-doped fiber amplifier (EDFA) offers an ideal broadband source with high brightness. Superluminescent diodes can alternatively be used and cover also the near-infrared with very similar properties. The experimental setup and measurements will be described in section \ref{setup}. It is important to remember that measurements in the spectral domain, where the fringe spacing is simply the free spectral range of the etalon, give the optical path length with the group index and not the refractive index. 
We will see in section \ref{new_sellmeir} how to use these measurements to extrapolate index values in the near-infrared.

To illustrate the interest of our accurate index measurement in section \ref{thin_film}, we will use it to evaluate the thickness of a thin film of \yso by collecting the transmission with a commercial moderate-resolution spectrometer.

%The paper is organized as follows. We first describe the experimental setup to probe an oriented sample in three orthogonal directions. We simply rotate the polarization to reveal the birefringence. We then detail the signal processing of the interferograms to measure the refractives index values. We finally reposition our work with respect to previous work and propose ..??

\section{Extrapolating previous measurements}\label{Extrapolating}
%It is important to begin by analysing the historical measurements that are still relevant, carried out by Beach {\it et al.} in 1990 \cite{sellmeier}.

\begin{figure}[htbp]
\centering
\includegraphics[width=.75\textwidth]{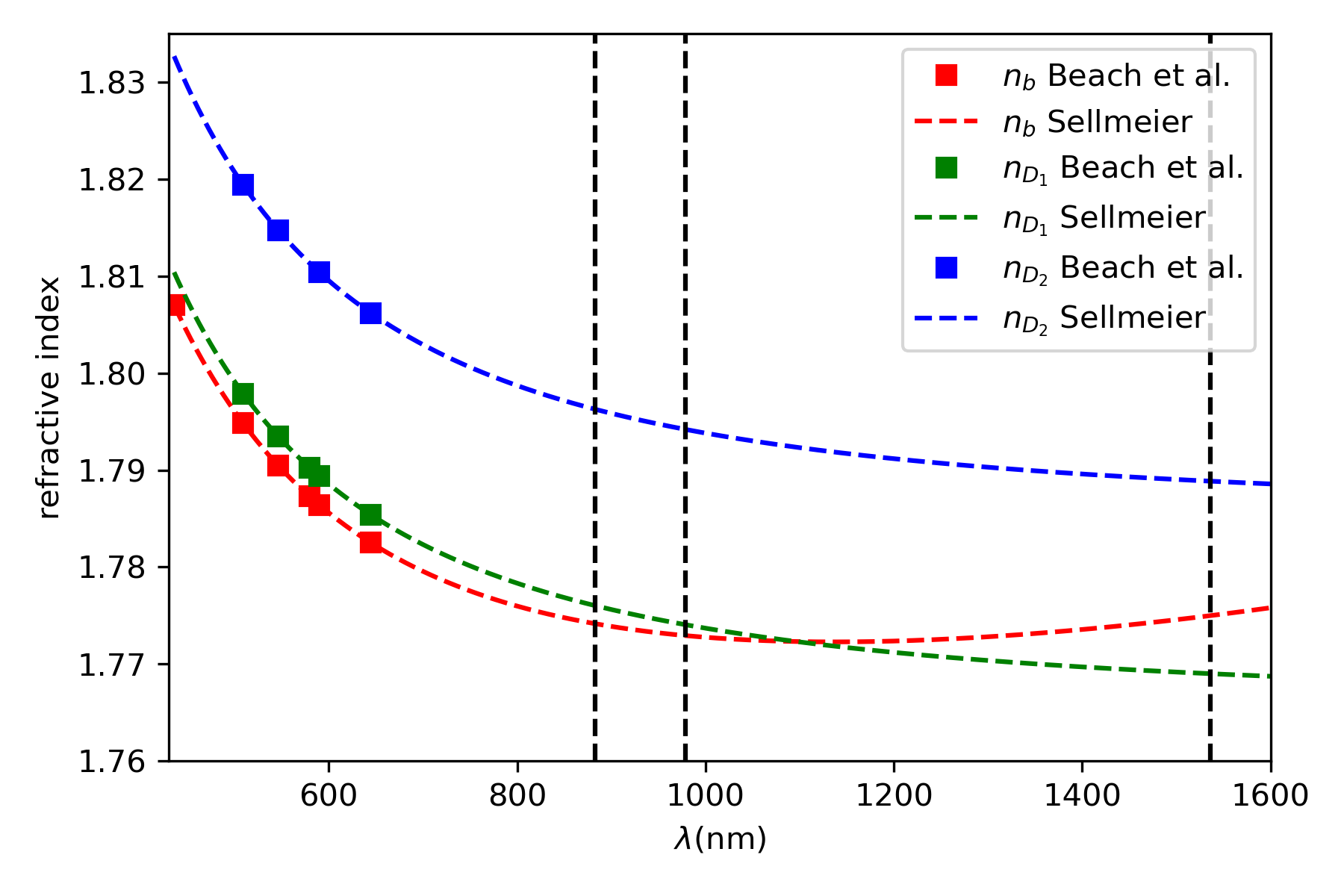}
\caption{Refractive index measurements by Beach {\it et al.} \cite{sellmeier} (squares) and the dispersion curves (Eq.\ref{eq:Sellmeier}) using the Sellmeier coefficients \ref{tab:Sellmeier} (dashed lines). We also represent the transitions of interest for neodymium, ytterbium and erbium at 883nm, 979nm and 1536 nm respectively.}
\label{fig:Beach_sellmeier}
\end{figure}

It is important to begin by analysing the historical measurements that are still used as a reference, carried out by Beach {\it et al.} in 1990 \cite{sellmeier}. They cover six wavelengths between 436nm and 644nm using spectral lamps (Hg, Cd, Na) (Fig.\ref{fig:Beach_sellmeier}) and enable the dependencies of the principal indices to be obtained using a Sellmeier formula whose validity we will discuss a little later. We reproduce it here in its original form and will use it throughout the rest of this paper.

\begin{equation}
n_i(\lambda)^2 = A_i + \frac{B_i}{\lambda^2+C_i}+D_i \lambda^2
\label{eq:Sellmeier}
\end{equation}

$ \lambda $ is the wavelength expressed in $\mu$m and $A_i$, $B_i$, $C_i$, $D_i$, the Sellmeier coefficients along the axis $i$ (where $i=$\bax, \Du\, or \Dd), as tabulated below:

\begin{table}[!h]
\begin{center}
\begin{tabular}{| c |c |c |c |c| }
\hline
axis $i$ &$ A_i$ & $B_i$ & $C_i$ & $D_i$ \\
 \hline
$n_b$ & 3.0895 & 0.0334 & 0.0043 & 0.0199 \\
\hline
$n_{D_1}$ & 3.1173 & 0.0283 & -0.0133 & 0.00 \\
\hline
$n_{D_2}$ & 3.1871 & 0.03022 &  -0.0138 & 0.00 \\
\hline
\end{tabular}
\end{center}
\caption{Sellmeier coefficients along the axes  \bax, \Du, \Dd\, from Beach {\it et al.} \cite{sellmeier}.}
\label{tab:Sellmeier}
\end{table}

Regardless of how accurately the parameters are fitted, it is important to discuss the physical significance of the equation \ref{eq:Sellmeier}. This is the two-pole
Sellmeier dispersion formula \cite{Ghosh:97}, whose general expression should be \begin{equation} n_i(\lambda)^2 = A^\prime_i + \frac{B^\prime_i}{\lambda^2-C^\prime_i}+\frac{B^{\prime\prime}_i}{\lambda^2-C^{\prime\prime}_i}
\end{equation}

with an ultraviolet pole (coefficients $B^\prime_i$ and $C^\prime_i$ with a resonant wavelength at $\sqrt{C^\prime_i} $) and a mid-infrared pole ($B^{\prime\prime}_i$ and $C^{\prime\prime}_i$ resonant at $\sqrt{C^{\prime\prime}_i} $) \footnote{ One also finds the following general expression, which is algebraically equivalent $n_i(\lambda)^2 = A^\prime_i + \frac{B^\prime_i \lambda^2}{\lambda^2-C^\prime_i}+\cdots$}. Assuming $\lambda$ far from mid-infrared resonance ($\lambda \ll \sqrt{C^{\prime\prime}_i}$), the expression can be simplified as
\begin{equation} n_i(\lambda)^2 = \left(A^\prime_i -\frac{B^{\prime\prime}_i}{{C^{\prime\prime}_i}} \right)+ \frac{B^\prime_i}{\lambda^2-C^\prime_i}-\frac{B^{\prime\prime}_i}{\left({C^{\prime\prime}_i}\right)^2} {\lambda^2}
\end{equation}

This last expression is identical to the expression \ref{eq:Sellmeier} used to define the coefficients $A_i$, $B_i$, $C_i$ and $D_i$. From a qualitative point of view, it should be noted that $C_i$ and $D_i$ must be negative to make sense of the ultraviolet and infrared resonances. We can already see from the table \ref{tab:Sellmeier} that this is not the case for $n_b$. This contradicts the assumptions of Sellmeier’s model. We can actually estimate $C_i$ given the position of the gap in \yso \cite{PANG20053539_gap}. With 6.14 eV, we expect $C_i \sim -0.04$, which gives only an order of magnitude \cite[eq.2]{Ghosh:97}. Regarding the mid-infrared pole, a single term in $D_i$ seems reasonable. We are indeed far from the resonances that appear around 1000 cm$^{-1}$, or 10000 nm associated to stretching of the Si-O-Si bond \cite[and references therein]{molecules30214161_FTIR}.

We plotted the various experimental values and dispersion curves based on tabulated coefficients extended to the optical telecommunications range, which includes the erbium transition at 1536 nm actively studied in the context of classical and quantum optical processing\footnote{corresponding to the so-called site 1 of yttrium substitution and between the lowest crystal levels} in Fig.\ref{fig:Beach_sellmeier}.

%/home/thierry/neel_ownCloud/neel_exchange/YSO_index/Sellmeier/

Extrapolation into the near-infrared, far from the measurement range, does not really make sense since there is a crossover of indices around 1100 nm. Even though the monoclinic structure essentially explains the anisotropic nature and differences in indices, a crossover of the indices is very unlikely. As $D_{b}$ is positive, the index $n_b$ increases, which is not physically plausible as already explained in the light of the Sellmeier model. Furthermore, this would induce, for example, a complete change of plane for the two optic axes of a crystal (direction with no birefringence), which vary very little (by a few degrees), as confirmed in monoclinic crystals similar to \yso \cite{Segonds:04}.
Taking a positive value of the $D_b$ Sellmeier coefficient lacks of physical meaning, unlike $D_{D_1}=0$ and $D_{D_2}=0$ (Table \ref{tab:Sellmeier}). This undoubtedly induces a better fit of the experimental measurements in the visible range at the cost of a non-physical branch in the infrared. To be more quantitative, the root mean square deviation (RMSD) error  for Beach {\it et al.}'s fit is 3.0$\times 10^{-5}$ and goes up if we additionally impose $D_b$=0 to 1.1$\times 10^{-3}$.

Before discussing the accuracy of the fits, we need to consider the possible sources of systematic error in index measurements.

First, there is an intrinsic limitations of the precision because of the monoclinic nature of \yso. The orientation of the dielectric reference frame (\bax, \Du, \Dd), which should not be confused with the crystal frame ({\bf a, b, c}), is not fixed, but rotates with wavelength, with only \bax\, remaining constant \cite{Traum:14}. It is conventionally defined relative to the reference \cite{li1992spectroscopic}, but it would be preferable to measure it at each wavelength. Strictly speaking, one would need to define the orientation of the reference frame and measure the values of the indices for each wavelength. This is not realistic. One therefore prefers to choose a fixed orientation for the reference frame—say, an average one—and model the indices using Sellmeier’s formula. This limits the precision with which the indices can be estimated. This intrinsic error can be evaluated as follows. As measured by Traum {\it et al.}\cite{Traum:14}, the reference frame  (\bax, \Du, \Dd) rotates by approximately 3$^{\circ}$ between the visible and near-infrared regions. Such a rotation is sufficient to introduce uncertainty into the refractive index: with a refractive index of approximately $\sim$ 1.8 and a 0.01 maximum index difference between \bax\,  and \Dd, the uncertainty will then be approximately 1$\times 10^{-4}$. We need to bear this order of magnitude in mind. Any estimate that surpasses this value and assumes a fixed reference frame will therefore be sufficient.

Second, as mentioned in the introduction, the historical measurements have been performed with a 10\%-doped \ndyso sample \cite{sellmeier} whose index can be significantly different than undoped or low doping crystals. It is difficult to find literature on the subject. Zelmon {\it et al.} observe an index change of approximately 3$\times 10^{-3}$ for an 8\%-ceramic \cite{Zelmon2005}. This order of magnitude is consistent with Soharab {\it et al.}’s measurements \cite{Soharab2019}. This gives an idea of the potential error if the doped crystal is taken as the starting point.

To conclude, our objective is to reconcile the values in the visible and near-infrared by performing an independent measurement. As we shall see, it is entirely possible to achieve sufficient accuracy (<1$\times 10^{-4}$), below the systematic errors, whilst ensuring that the Sellmeier coefficients retain a physical meaning ($C_i$ and $D_i$ negative).

\section{Experimental measurements}\label{setup}

\subsection{Experimental setup}
We use a \yso crystal from Scientific Materials, slightly doped with 50 ppm of \er, as a Fabry-Perot etalon. The crystal faces are sufficiently reflecting with a typical Fresnel reflexion coefficient of 8\% (index $\sim$ 1.8) to observe fringes with a spectrometer.

At our level, we consider it as {\it pure} \yso because \er is only present at the level of traces. In that sense, the impurities do not significantly distort the crystal cell and in turn the refractive index as it can be expected for the \ndyso sample used by Beach {\it et al.} with 10\% doping \cite{sellmeier}. The measurement are performed at room temperature where the narrow absorption line of \er can be neglected at this doping level and temperature. \yso is monoclinic where the crystal \bax-axis, the special monoclinic axis, defines a reference of the dielectric frame. The two other principal axes lie in the ({\bf a,c}) plane and are usually called \Du and \Dd. The orientation of the dielectric frame (\bax, \Du, \Dd) is defined in \cite[Fig.1.(iii)]{PETIT2020100062} consistently with \cite{li1992spectroscopic} and is realized experimentally with a Laue diffractometer (sub degree precision). The facets parallelism is rectified by lapping and polishing with the same precision. The final dimensions of the all-side-polished crystals are 3.04, 4.86 and 5.79 mm along \bax, \Du\, and \Dd\, respectively.

\begin{figure}[htbp]
\centering
\includegraphics[width=.7\textwidth]{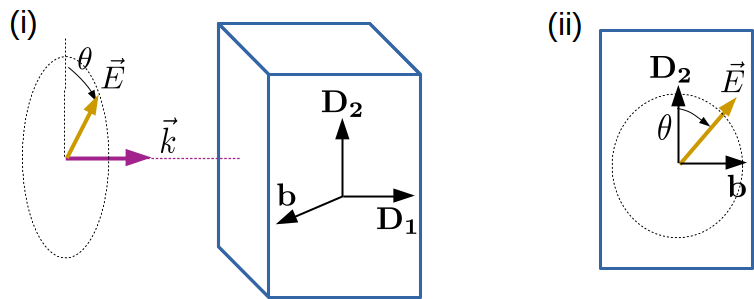}
\caption{(i) The \yso crystal is all-side-polished. We probe the birefringence by rotating the polarization by an angle $\theta$. In this example, the light propagates along \Du\, and the index is measured in the (\bax,\Dd) plan as a function of  $\theta$. We reproduce the experiment along \bax\, and \Dd\, as propagation axis. }
\label{fig:cuboid}
\end{figure}

Our broad-band light source is an EDFA delivering typically 10\,mW out of a single mode fiber and covers 40nm around 1550\,nm. We collimate the output of the fibered source (beam size $\sim$ 1mm) and align it through the sample. The transmission of the crystalline etalon is fiber-coupled and recorded by an optical spectrum analyzer Ando AQ6317B (OSA) whose resolution is 0.01\,nm. We insert a polarizer and a half-wave plate (HWP) to control the light polarization impinging the crystal and we rotate the HWP to study the birefringence of \yso. An example of the dielectric frame orientation with the definition of the propagation axis and light polarisation is represented in Fig.\ref{fig:cuboid}.

\begin{figure}[htbp]
\centering
\includegraphics[width=.75\textwidth]{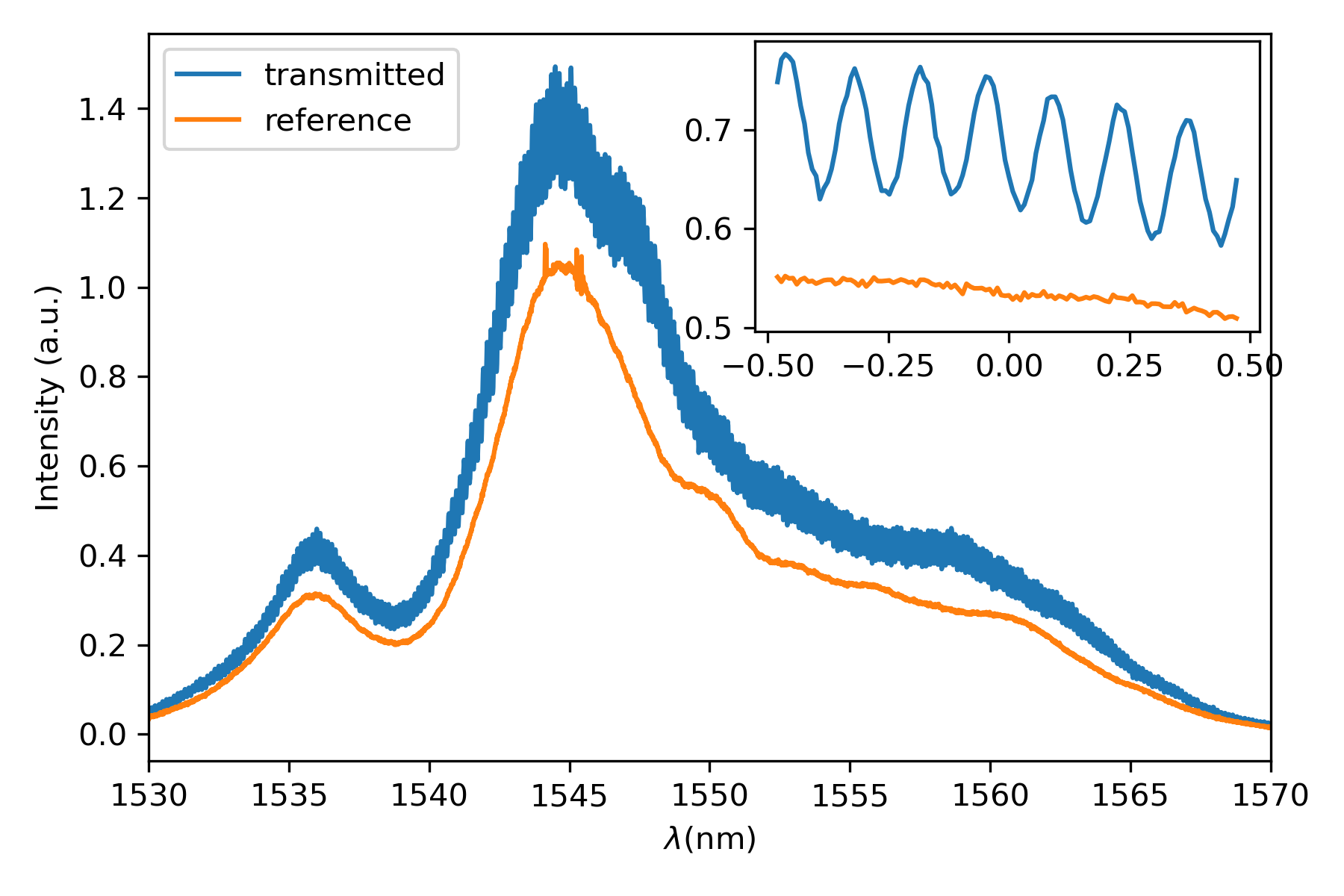}
\caption{Spectrum of the transmitted light propagating along \Du\, (blue curve). The incoming polarization is parallel to \Dd\, corresponding to Fig.\ref{fig:cuboid} with $\theta=0$. A reference spectrum is recorded without the crystal in the path to characterize the light source (orange curve). Inset: Zoom on the central part of the interferogram (centered on 1550\,nm and 1\,nm-broad) where interference fringes are clearly visible ($\sim$ 0.14\,nm spacing).}
\label{fig:Trait_k_5mm_HW0deg}
%17:58:45: File /home/thierry/neel_ownCloud/20211004_YSOindex/Trait_k_6mm_HW0deg.py saved.
\end{figure}

We show an example of recorded optical spectrum in Fig.\ref{fig:Trait_k_5mm_HW0deg} where Fabry-Perot interferences are visible. 

The fringe spacing in Fig.\ref{fig:Trait_k_5mm_HW0deg}, namely the free-spectral-range (FSR), is roughly 0.14\,nm for a $L_\mathrm{D1}$=4.86\,mm propagation in \yso. This confirms that the refractive index is is close 1.8 \cite{sellmeier, PhysRevB.67.245108}. A detailed analysis of the interferograms, when we rotate the polarization, allows us to extract accurately the principal values of the index in the different directions.

\subsection{Fitting the group index}

The FSR of the observed modulation $\Delta_\mathrm{FSR}$ allows the optical length to be obtained, which is the product of the physical length and the index. It is important to remind here that this measures the group index and not the refractive index directly. By definition, when the frequency $\nu$ varies by $\Delta_\mathrm{FSR}$, then the optical phase $k(\nu)L$ varies by 2$\pi$ ($k$ is the wavevector) i.e. $2 k(\nu+\Delta_\mathrm{FSR})L-2 k(\nu)L = 2\pi$. This yields to 

\begin{equation}
\Delta_\mathrm{FSR} = \frac{c}{2 \mathcal{N} L}
\label{eq:FSR}
\end{equation}
where $\mathcal{N}=\displaystyle\frac{c}{2\pi} \frac{\partial k }{\partial \nu} $ is the group index that can be also written as  
\begin{equation}\mathcal{N}=\displaystyle\frac{c}{2\pi} \frac{\partial k }{\partial \nu}=n-\displaystyle\lambda \frac{\partial n }{ \partial \lambda}
\label{eq:n_group}
\end{equation}
where $n$ is the refractive index and $ \lambda$ the vacuum wavelength.
To extract the oscillation period, the FSR $\Delta_\mathrm{FSR}$, we will perform a Fourier transform of the interferograms, whose Fig.\ref{fig:Trait_k_5mm_HW0deg} is an example, noted $\tilde{T}$, and defined as $\tilde{T}(\tau)  = \displaystyle \int_{\nu} T(\nu)\, \exp({-2\mathrm{i}\pi \, \tau  \nu}) \,\mathrm{d} \nu$.

It can be plotted directly in units of the group index $\mathcal{N}$ by calculating
\begin{equation}
\tilde{T}(\mathcal{N})  = \int_{\nu} T(\nu)\, \exp({-4 \mathrm{i}\pi \, \mathcal{N} L \, \nu / c}) \,\mathrm{d} \nu
\label{eq:Fourier}
\end{equation}
whose absolute value exhibits two peaks under birefringence, for example at $\mathcal{N}_{b} $ and $\mathcal{N}_{D_2}$ when propagating along \Du.

\begin{figure}[htbp]
\centering
\includegraphics[width=.95\columnwidth]{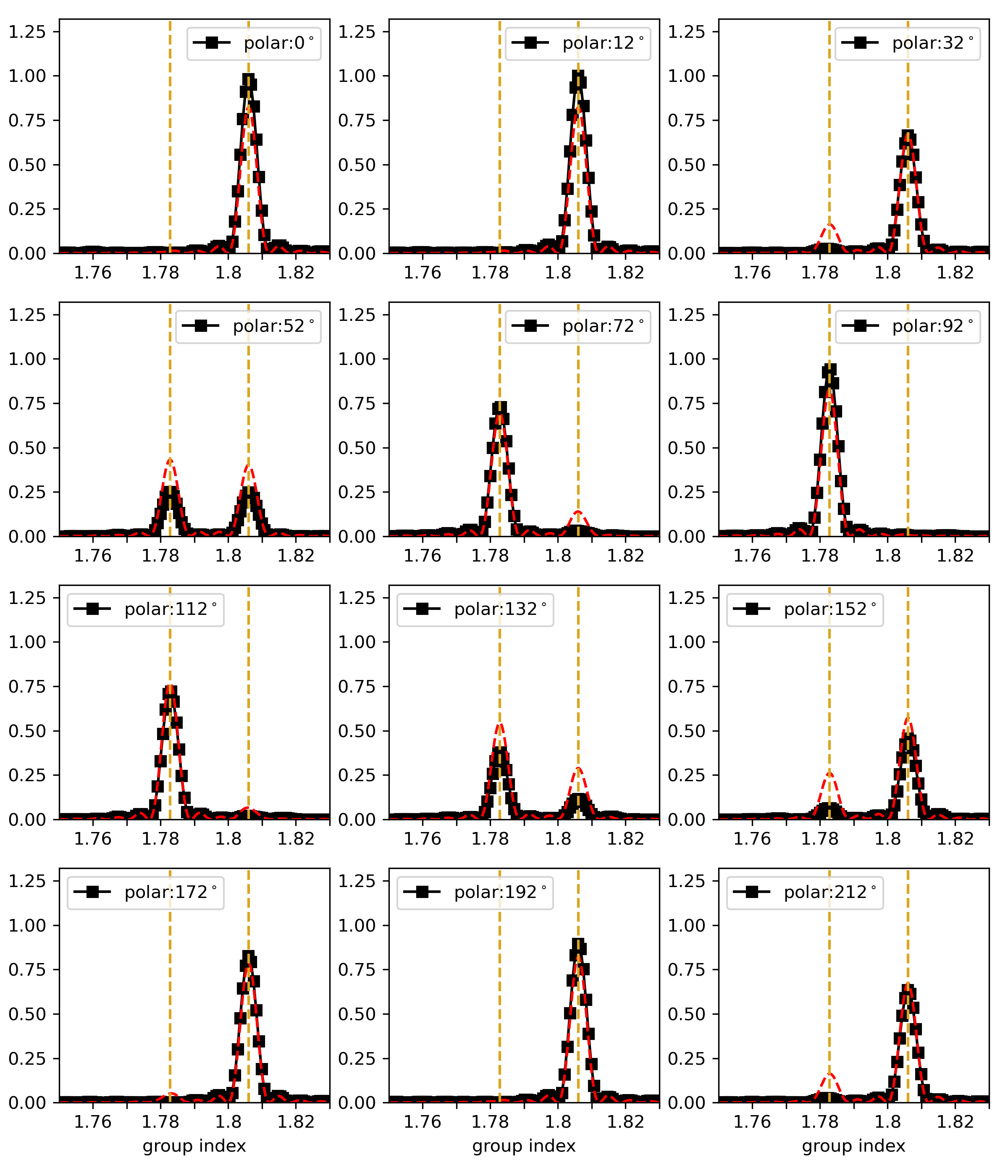} 
\caption{Absolute value of $\tilde{T}(\mathcal{N})$ (Eq. \ref{eq:Fourier} in abritrary units) when propagating along \Du\,. The two peaks marked with vertical dashed lines reveal the birefringence. The values of $\mathcal{N}_{b} $ and $\mathcal{N}_{D_2}$ can be extrated from a global processing of the polarisation rotation pattern.}
\label{fig:Fourier_spectra_k_5mm}
%/home/thierry/neel_ownCloud/neel_exchange/YSO_index/20211004_YSOindex/trait_OSA_fit_global_sinc.py
\end{figure}

This can be seen in the Fig.\ref{fig:Fourier_spectra_k_5mm} when we rotate the polarisation. Zero degree corresponds to polarisation along \bax\, with a peak around 1.785 and 90$^\circ$ corresponds to \Dd\,\footnote{The experimental polarisation is more precisely 92$^\circ$, which depends on the reference position at 0$^\circ$ and the fixed steps of the waveplate motorised rotation stage.} with a peak around 1.805. Between these two positions, two peaks can be observed, illustrating the birefringence. Processing these curves, which we will now describe, allows us to extract precise values for both indices, $\mathcal{N}_{b} $ and $\mathcal{N}_{D_2}$ in that case.

The plots in Fig.\ref{fig:Fourier_spectra_k_5mm} exhibit two peaks whose shape is a $\mathrm{sinc}^2$ function and whose linewidth $\Delta_\mathcal{N}$ is limited by the total frequency span of the interferogram namely $\Delta_\nu=$4992.2 GHz corresponding to the range 1530-1570\,nm as in Fig. \ref{fig:Trait_k_5mm_HW0deg}. Since we choose to plot $\displaystyle \tilde{T}$ in units of $\mathcal{N}$, the linewidth is $\Delta_\mathcal{N}=\displaystyle \frac{c}{2 L \Delta_\nu}$. The elementary fitting function will be $\displaystyle \mathrm{sinc}^2\left(\frac{\mathcal{N}-\mathcal{N}_i}{\Delta_\mathcal{N}} \right)$ with a fixed width $\Delta_\mathcal{N}$ and $\mathcal{N}_i$ a free fitting parameter.

For each curve in the Fig.\ref{fig:Fourier_spectra_k_5mm}, the two peaks noted $\mathcal{N}_1$ and $\mathcal{N}_2$ are identically positioned and their relative weights correspond to the projection of the incident polarisation onto the principal axes which a known experimental parameter. Then, we can fit all the curves at once, since they are a composition of sinc functions between two indices $\mathcal{N}_1$ and $\mathcal{N}_2$ that we seek to estimate, and whose amplitudes depends on the known angle of polarisation $\theta$. The global fitting function is then

\begin{equation}
 \cos^2(\theta-\theta_0)\, \mathrm{sinc}^2\left(\frac{\mathcal{N}-\mathcal{N}_1}{\Delta_\mathcal{N}} \right) +  \sin^2(\theta-\theta_0)\, \mathrm{sinc}^2\left(\frac{\mathcal{N}-\mathcal{N}_2}{\Delta_\mathcal{N}} \right)
 \label{eq:sinc2}
\end{equation}

where we add $\theta_0$ to compensate for an initial imperfect alignement of the polarisation with respect to one of the principal axes. There is also a global scaling amplitude parameter since we plot $\tilde{T}(\mathcal{N})$ in arbitrary units. With $\mathcal{N}_1$ and $\mathcal{N}_2$, representing alternatively  $\mathcal{N}_{b} $, $\mathcal{N}_{D_1}$ and $\mathcal{N}_{D_2}$ depending on the axis of propagation, there are only 4 fitting parameters ($\theta_0$, a global amplitude and the two indices of interest) for the 12 curves for different $\theta$.

To lighten the core of the article somewhat, we have only shown the propagation according to \Du\,  in the Fig.\ref{fig:Fourier_spectra_k_5mm} and added the curves corresponding to the propagation along \bax\, and \Dd\, in the appendix \ref{appendix:patterns}.

\subsection{Principal values of the index}\label{sec:group_indices}

The fitted value are summarized in Table \ref{tab:group_indices}. There is redundancy between the values since the same index can be extracted with two different propagation axes.

%%%ampl,n1,n2,theta_0_deg

%%---k_6mm--- D2
%%nu span:4992.2GHz
%%f value init:0.023
%%f value final:0.007
%%optim - ampl,n1,n2,theta_0_deg:[ 0.79507583  1.78583443  1.78885719 72.17993656]
%%optim - n1,n2:1.78583+-0.0095,1.78886+-0.0074
%%---k_5mm--- D1
%%nu span:4992.2GHz
%%f value init:0.021
%%f value final:0.009
%%optim - ampl,n1,n2,theta_0_deg:[ 0.83097419  1.78280672  1.80604912 71.9862847 ]
%%optim - n1,n2:1.78281+-0.0134,1.80605+-0.0088
%%---k_3mm--- b axis
%%nu span:4992.2GHz
%%f value init:0.030
%%f value final:0.011
%%optim - ampl,n1,n2,theta_0_deg:[ 0.79621807  1.78697712  1.80733787 72.04852636]
%%optim - n1,n2:1.78698+-0.0202,1.80734+-0.0128

\begin{table}[!h]
\begin{center}
\begin{tabular}{| c |c |c |c | }
\hline
Group indices & \bax\, propagation & \Du\, propagation & \Dd\, propagation  \\
 \hline
$\mathcal{N}_b$ &  & 1.783$\pm$0.0134 & 1.786$\pm$0.0095 \\
\hline
$\mathcal{N}_{D_1}$ & 1.787$\pm$0.0202 &  & 1.789$\pm$0.0074 \\
\hline
$\mathcal{N}_{D_2}$ & 1.807$\pm$0.0128 & 1.806$\pm$0.0088 &  \\
\hline
\end{tabular}
\end{center}
\caption{Group indices $\mathcal{N}_b$, $\mathcal{N}_{D_1}$ and $\mathcal{N}_{D_2}$ extracted from Fig.\ref{fig:Fourier_spectra_k_3mm} (\bax\, propagation), Fig.\ref{fig:Fourier_spectra_k_5mm} (\Du\, propagation)  and Fig.\ref{fig:Fourier_spectra_k_6mm} (\Dd\, propagation)  resp. }
\label{tab:group_indices}
\end{table}

The measurements are generally consistent, and the two measured values of $\mathcal{N}_b$, for example, are compatible within their error bars. These error bars are of the order of $10^{-2}$, as given by the covariance matrix during the fitting procedure. In any case, it is not possible to eliminate the systematic error in the measurement of the propagation length $L$ (0.01\,mm), which is $3\times10^{-3}$ for $L=3.04$\,mm.

For the rest, we decided to simply keep the average of the two measurements for each axis, i.e. the group index values $\mathcal{N}_b$ =1.784, $\mathcal{N}_{D_1}$=1.788 and $\mathcal{N}_{D_2}$=1.807.

%\section{Signal processing of the interferograms}

\section{Extending the Sellmeier coefficients in the near-infrared}\label{new_sellmeir}

These values allow us to extend the index measurements in the infrared and thus modify the historical Sellmeier coefficients. It is not possible to use the group index measurements as they are, since we are seeking to predict the refractive index. However, we can reasonably assume that the difference between the group and refractive indices $\mathcal{N}_{i} -n_i$ for a given wavelength (1550\,nm in our case) is independent of the axes $i$ (where $i=$\bax, \Du\, or \Dd). This is equivalent to assuming in eq.\ref{eq:n_group} that the dispersion term $\displaystyle \frac{\partial n }{ \partial \lambda}$ is independent of the axis. 
This is clearly a simplifying assumption, the validity of which is not easy to verify a priori. For example, it would need to be tested on a similar crystal, given that the bandgap of \yso is particularly large, and using data across a wide range of wavelengths, which we have not been able to identify. This is a point to bear in mind for future studies.
%From a microscopic point of view, the dispersion characterised by Sellmeier's equation can be seen as a manifestation, far from resonance, of an electronic phenomenon linked to the atoms themselves that compose the matrix. We therefore expect this to depend little on the arrangement of atoms along the different axes. We have already used this argument to explain the non-physical nature of an index crossover as a function of wavelength in \ref{Extrapolating}. Here, we go further by assuming that $\mathcal{N}_{i} -n_i$ is independent of the axes. [BENOIT?]

In very practical terms, to extend the fit, we do not use the index measurements directly, but rather their differences, assuming that $\mathcal{N}_{D_2}-\mathcal{N}_{D_1}=n_{D_2}-n_{D_1}$ and $\mathcal{N}_{D_1}-\mathcal{N}_{b}=n_{D_1}-n_{b}$. We therefore take these two differences in indices as new data for adjusting the coefficients.

Furthermore, as we discussed in \ref{Extrapolating}, introducing the Sellmeier coefficients $D_i$ into the equation may cause the index to increase at long wavelengths especially if it is improperly choosen positive. We therefore decided to impose $D_i = 0$ as well. This is a safe assumption all the more so as we are still far from mid-infrared poles ($\sim$ 10000 nm). We will see that the accuracy achieved is more than adequate. To complement our analysis and cover the topic in full, we examine in the appendix \ref{appendix:Dnegative} the case where $D_i < 0$ is a non-zero negative.

\begin{figure}[htbp]
\centering
\includegraphics[width=.75\textwidth]{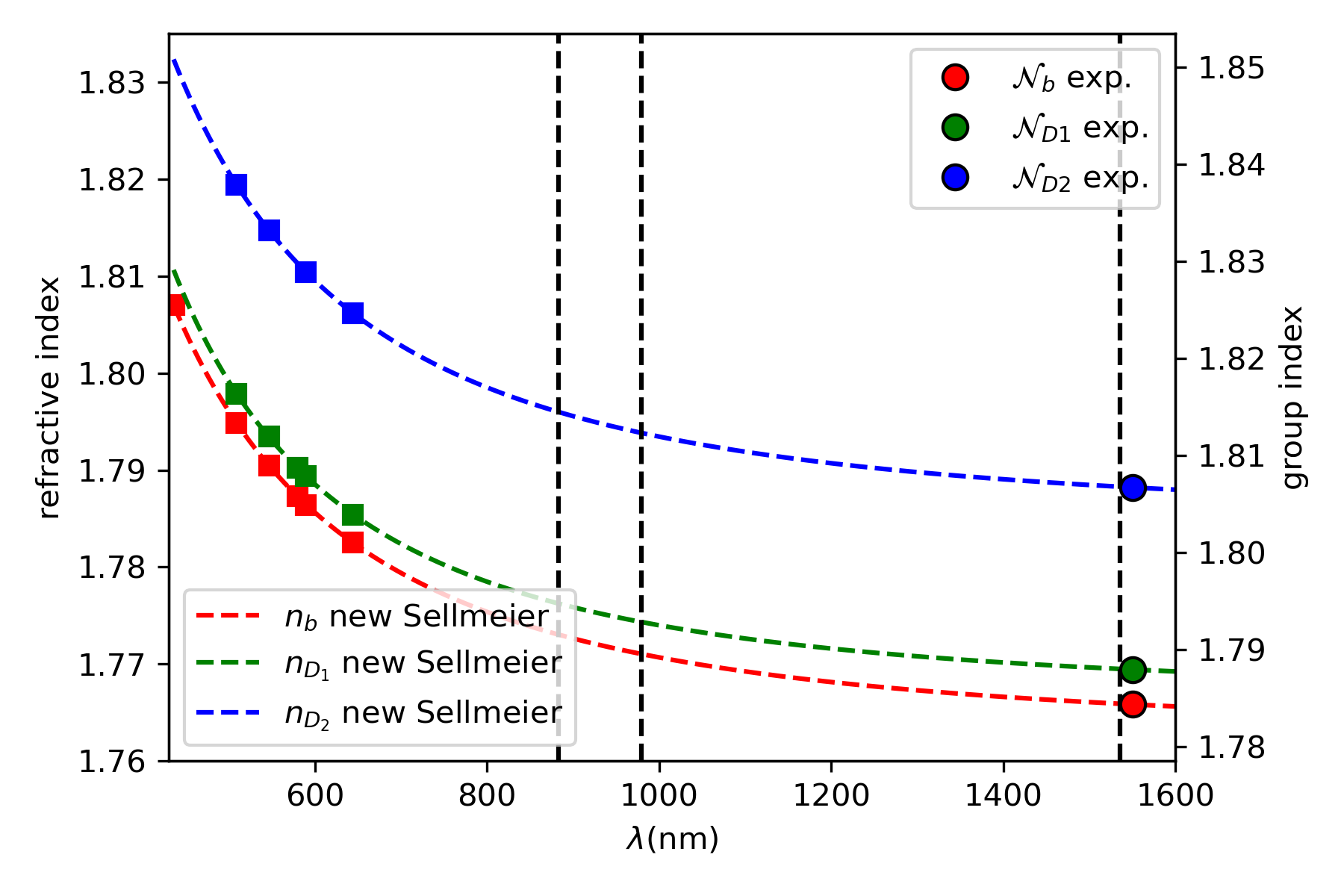}
\caption{Extended Sellmeier coefficients in the near-infrared using the group index measurements described in \ref{sec:group_indices}. Our measurements $\mathcal{N}_b$, $\mathcal{N}_{D_1}$ and $\mathcal{N}_{D_2}$ are represented with circles corresponding to the right scale. See text for details.}
\label{fig:Neel_sellmeier}
\end{figure}

So we take the historical measurements (markers in Fig.\ref{fig:Beach_sellmeier} from \cite{sellmeier}), add the two index differences measured by us, and set $D_i$ to zero.  
The coefficients $A_i$, $B_i$ and $C_i$ are fit by using the least squares method (minimizing the RMSD error). The result is shown in the Fig. \ref{fig:Neel_sellmeier}.

%/home/thierry/neel_ownCloud/neel_exchange/YSO_index/Sellmeier/

The RMSD error is 3.2$\times 10^{-5}$, not very different from that of Beach {\it et al.} (3.0$\times 10^{-5}$). That is sufficient in any case, as we discussed at the end of section \ref{Extrapolating}. It is certainly reassuring to find negative values for $C_i$ to make sense of the ultraviolet pole.

We also extract the constant difference between the group and the refractive index $\mathcal{N}_{i} -n_i$ = 0.018 at 1550\,nm, which allows us to represent our measurements with a shift on the same figure \ref{fig:Neel_sellmeier} (on the left scale with the refractive index, and on the right scale with group index corresponding to our measurements in Table \ref{tab:group_indices}).
The coefficients corresponding to the three Sellmeier equations are summarised in the following Table \ref{tab:Sellmeier_new}:

%----optim with D=0 and Neel meas.-----
%[ 3.1060  3.1196  3.1844  0.0290  0.0269  0.0319 -0.0076 -0.0205 -0.0057]
%RMSD final Neel:3.2e-05

\begin{table}[htbp]
\begin{center}
\begin{tabular}{| c |c |c |c |c| }
\hline
axis $i$ &$ A_i$ & $B_i$ & $C_i$ & $D_i$ \\
 \hline
$n_b$ &3.1060 & 0.0290  & -0.0076  & 0.00 \\
\hline
$n_{D_1}$ & 3.1196 & 0.0269  & -0.0205 & 0.00 \\
\hline
$n_{D_2}$ & 3.1844 & 0.0319 & -0.0057& 0.00 \\
\hline
\end{tabular}
\end{center}
\caption{Extended Sellmeier coefficients in the near-infrared. We neglect the influence of mid-infrared poles by setting $D_i=0$ (see text for details)}
\label{tab:Sellmeier_new}
\end{table}
We propose that these coefficients be used henceforth to model the \yso index in both the visible and near-infrared ranges covering most of the rare-earth dopants.

\section{Application to thin film thickness measurement} \label{thin_film}
As an illustration, we propose here a way of turning the problem around and using the interferometric method described in section \ref{setup} to measure the thickness of a thin film (typically 1-10$\mu$m). Since the index is known, measuring the optical path gives the physical length. Conversely, starting from a calibrated thickness as we did in section \ref{setup}, we obtained the index.

For thin films, the constraint on the resolution of the spectrometer is much less limiting. Here we use a device that is fairly common in laboratories, a transmission spectrometer Perkin Elmer Lambda 900 UV-Vis-NIR.

\begin{figure}[htbp]
\centering
\includegraphics[width=.95\columnwidth]{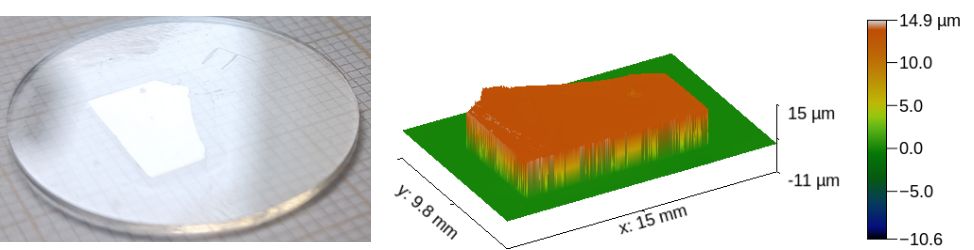} 
\caption{Left: picture of  thin \yso of glass. Right: Optical profilometric profile of the fim. }
\label{fig:YSO14um}
%/home/thierry/sDrive_CNRS/Innovation_board/plan_quantique/PEPR_MémoiresQ/AG_QCOMMTESTBEDxQMEMOxDIQKD_13mars2026/YSO_14um/
\end{figure}

The thin \yso sample is obtained as follows. A 10x10x1 mm plate is first polished to a flatness of lambda/10 and a roughness of 0.8\, nm, then directly bonded to a borosilicate glass substrate of equivalent flatness by so-called optical contacting. The crystal adhered to the substrate is then lapped to a target thickness of 10-20 $\mu$m and then polished again. Partial bonding in obtained for a surface of 10x5 mm as observed in Fig.\ref{fig:YSO14um}. The unbonded part is then stripped away during the lapping process.

It is difficult to extract the exact thickness of the step profile on the sides of the film using the Bruker ContourGT-K optical profilometer in our case (Fig.\ref{fig:YSO14um}, right). It is typically 14\,$\mu$m depending on the scanning region. It is therefore particularly useful to have an optical see-through measurement that is sensitive to the refractive index of \yso.

\begin{figure}[htbp]
\centering
\includegraphics[width=.75\columnwidth]{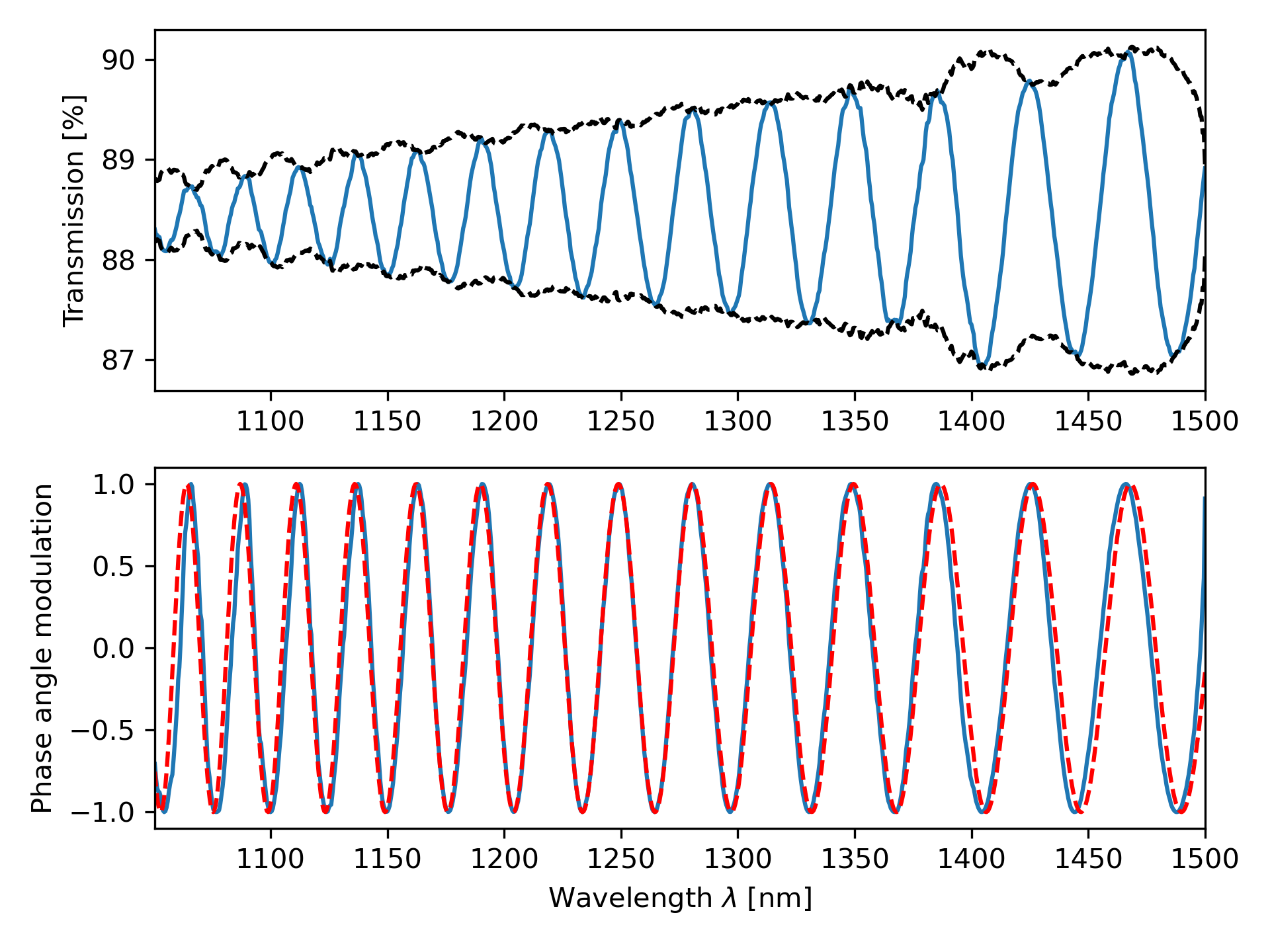} 
\caption{Top: transmission spectrum of a thin \yso film bonded to a glass substrate. The light is polarized along the \bax axis. Bottom: The oscillating part (noted Phase angle modulation) is renormalised by computing the Hilbert transform and dividing by the enveloppe (dashed lines in the top panel). The red dashed line is a fit to data to obtain the sample thickness using the know value of the index along \bax\, (Table \ref{tab:Sellmeier_new}).}
\label{fig:fit_fringes}
%/home/thierry/neel_ownCloud/QMEMO_cloud/20251222_Scan-Lambda950/
\end{figure}

The \bax\, axis will serve as our reference; it is first identified by Laue diffraction. The light from the spectrometer is then polarised along this axis. The white-light beam size delimited by a diaphragm of 4 millimeters which covers well the film center. We therefore record the transmission in the near-infrared where the spacing between fringes is greatest, as can be seen in the figure \ref{fig:fit_fringes} (top).

Fringes are clearly observed. We extract the oscillating part by computing the Hilbert transform and by dividing by the envelope (Fig.\ref{fig:fit_fringes}, bottom). Thus renormalised, the expression of the fringes can be adjusted simply by the function
\begin{equation}
\cos\left( 2 \pi \frac{2n_b(\lambda) L_f}{\lambda}\right)
 \label{eq:cos}
\end{equation}
 where the length of the film $L_f$ is the only adjustable parameter since $n_b(\lambda)$ is given by eq.(\ref{eq:Sellmeier}) with the coefficients of Table \ref{tab:Sellmeier_new}. The fit gives a value of $L_f = 14.13\, \mu$m (red dashed line in Fig. \ref{fig:fit_fringes}, bottom). 
The fit error given by the covariance matrix is on the order of nanometer. Furthermore, it should be remembered that the precision of the index measurement is  3.2$\times 10^{-5}$ which gives the same order of magnitude. However, the flatness of the surface is on the order of lambda/10, which here represents a few tens of nm. We therefore only keep 4 significant digits.

\section{Conclusion}

%A partir de mesure interferométric indépendante, nous étendons la modélisation de l'indice d'YSO dans l'infrarouge. La précision des mesures est intrinsequement limité par la caractère monoclinique du cristal qui reste le cas le plus complexe en optique cristallin. Nous avons profité de cet article pour rappeler ces éléments et discuté du modèle de Sellmeier qui ne se résume pas une série de coefficients phénoménologiques, mais traduit l'existance de résonances, ultraviolet ou infrarouge, dont il est bon de connaitre au moins vaguement les positions.
%Les voies d'amélioration de ce travail sont nombreuses, mais demandent un effort expérimental important. Des mesures de réfracometrie à haute précision semblent difficiles inaccessibles. En effet, elles demandent trois échantillons suivant les axes diélectriques pour chaque longueur d'onde d'analyse puisque le repère diélectrique tourne légerement. Dans la mesure où nous disposons deja d'un modèle, on pourrait imaginer se limiter à trois longueurs d'onde par exemple (9 échantillons) et affiner les coefficients $A_i$, $B_i$ et $C_i$ suivant les trois axes, soit 9 paramètres également. Nous avons montré que la précision intriseque du fit était de quelques $\times 10^{-5}$ ce qui donnerait deja une bonne sur l'indice avec des échantillons parfaitement orientés.

Using independent interferometric measurements, we extend the modelling of the \yso refractive index in the near-infrared. The accuracy of the measurements is intrinsically limited by the monoclinic nature of the crystal, which remains the most complex case in crystal optics. We have taken the opportunity in this article to review these points and discuss Sellmeier’s model, which is not merely a set of phenomenological coefficients, but reflects the existence of resonances—in the ultraviolet or mid-infrared—whose positions it is useful to know at least approximately.
There are many avenues for improving this work, but they require significant experimental effort. High-precision refractometric measurements appear difficult to achieve. Indeed, they require three samples aligned along the dielectric axes for each analysis wavelength, since the dielectric reference frame rotates slightly. Since we already have a model, one could envisage limiting the analysis to three wavelengths, for example (9 samples), and refining the coefficients $A_i$, $B_i$ and $C_i$ along the three axes, giving 9 parameters in total. We have shown that the intrinsic accuracy of the fit is of the order of a few $\times 10^{-5}$, which would already provide a good estimate of the index with perfectly oriented samples.

%Il serait aussi interessant de compléter les mesures d'indices par des mesures d'absorption, les étant reliées par les relations de Kramers-Kronig, dans l' ultraviolet et l'infrarouge lointain (10000 nm), proches des résonances notamenent. On butera cependant à nouveau sur la question de l'orientation du repère diélectrique qui devrait avoir très fortement tourné entre ces deux régions extremes du spectre.

%YSO questionne de facon extreme les methodes de refractométrie, nous montrons cependant qu'il est possible obtenir un modèle simplifié suffisament précis pour couvrir les applications en technologie quantique. En tant que méthode, nous avons pris soin de clarifier l'interet de mesures interférométrique, qui donne l'indice de groupe, et de montrer comment elles peuvent intégrer dans un modèle de Sellmeier. L'illustration à la mesure d'épaisseur de couche mince correspond à une véritable attente dans la perspective d'intégration des systèmes actifs.

It would also be interesting to supplement the refractive index measurements with absorption measurements—the two being linked by the Kramers–Kronig relations—in the ultraviolet and mid-infrared (10000 nm) regions, particularly near the resonances. However, we will once again encounter the issue of the orientation of the dielectric reference frame, which is expected to have rotated significantly between these two opposite regions of the spectrum.

\yso poses extreme challenges for refractometry methods; however, we demonstrate that it is possible to obtain a simplified model that is sufficiently accurate to cover applications in quantum technology. As a method, we have taken care to clarify the interest of interferometric measurements, which provide the group index, and to show how they can be integrated into a Sellmeier model. The illustration of thin-film thickness measurement corresponds to a genuine expectation in the perspective of active material integration \cite{Thin_Films}.

\begin{backmatter}
\bmsection{Funding}
The author acknowledge support from the Plan France 2030 project QMEMO No. ANR-22-PETQ-0010, the project No. CHORIZO ANR-24-CE47-1190 and the Program QuanTEdu-France No. ANR-22-CMAS-0001 France 2030.

\bmsection{Acknowledgment}
We would like to thank R. Bouland for his assistance at the polishing workshop.

\bmsection{Disclosures}
The author declares no conflicts of interest.

\bmsection{Data availability}
Data underlying the results presented in this paper are not publicly available at this time but may be obtained from the authors upon reasonable request.

%\bmsection{Supplemental document}
%See Supplemental document for supporting content.

\end{backmatter}

\bibliography{YSO_index_bib}

@ARTICLE{sellmeier, author={R. {Beach} and M. D. {Shinn} and L. {Davis} and R. W. {Solarz} and W. F. {Krupke}}, journal={IEEE Journal of Quantum Electronics}, title={Optical absorption and stimulated emission of neodymium in yttrium orthosilicate}, year={1990}, volume={26}, number={8}, pages={1405-1412},}

@article{TAN1998158,
title = {Determination of refractive index of silica glass for infrared wavelengths by IR spectroscopy},
journal = {Journal of Non-Crystalline Solids},
volume = {223},
number = {1},
pages = {158-163},
year = {1998},
issn = {0022-3093},
doi = {https://doi.org/10.1016/S0022-3093(97)00438-9},
url = {https://www.sciencedirect.com/science/article/pii/S0022309397004389},
author = {C.Z. Tan}
}

@article{PhysRevB.67.245108,
  title = {Electronic and optical properties of {${\mathrm{Y}}_{2}{\mathrm{SiO}}_{5}$} and {${\mathrm{Y}}_{2}{\mathrm{Si}}_{2}{\mathrm{O}}_{7}$} with comparisons to \ensuremath{\alpha}-{${\mathrm{SiO}}_{2}$} and {${\mathrm{Y}}_{2}{\mathrm{O}}_{3}$}},
  author = {Ching, W. Y. and Ouyang, Lizhi and Xu, Yong-Nian},
  journal = {Phys. Rev. B},
  volume = {67},
  issue = {24},
  pages = {245108},
  numpages = {8},
  year = {2003},
  month = {Jun},
  publisher = {American Physical Society},
  doi = {10.1103/PhysRevB.67.245108},
  url = {https://link.aps.org/doi/10.1103/PhysRevB.67.245108}
}

@article{Singh_2002,
	doi = {10.1238/physica.regular.065a00167},
	url = {https://doi.org/10.1238/physica.regular.065a00167},
	year = 2002,
	month = {jan},
	publisher = {{IOP} Publishing},
	volume = {65},
	number = {2},
	pages = {167--180},
	author = {Shyam Singh},
	title = {Refractive Index Measurement and its Applications},
	journal = {Physica Scripta}
}

@article{SAINZ1994381,
title = {Real time interferometric measurements of dispersion curves},
journal = {Optics Communications},
volume = {110},
number = {3},
pages = {381-390},
year = {1994},
issn = {0030-4018},
doi = {https://doi.org/10.1016/0030-4018(94)90442-1},
url = {https://www.sciencedirect.com/science/article/pii/0030401894904421},
author = {C. Sáinz and P. Jourdian and R. Escalona and J. Calatroni}
}

@article{Galli:03,
author = {Matteo Galli and Franco Marabelli and Giorgio Guizzetti},
journal = {Appl. Opt.},
number = {19},
pages = {3910--3914},
publisher = {OSA},
title = {Direct measurement of refractive-index dispersion of transparent media by white-light interferometry},
volume = {42},
month = {Jul},
year = {2003},
url = {http://www.osapublishing.org/ao/abstract.cfm?URI=ao-42-19-3910},
doi = {10.1364/AO.42.003910},
}

@inproceedings{10.1117/12.439193,
author = {Di Yang and Michael E. Thomas and Simon G. Kaplan},
title = {{Measurement of the infrared refractive index of sapphire as a function of temperature}},
volume = {4375},
booktitle = {Window and Dome Technologies and Materials VII},
editor = {Randal W. Tustison},
organization = {International Society for Optics and Photonics},
publisher = {SPIE},
pages = {53 -- 63},
year = {2001},
doi = {10.1117/12.439193},
URL = {https://doi.org/10.1117/12.439193}
}

@article{Brindza:14,
author = {Michael Brindza and Richard A. Flynn and James S. Shirk and G. Beadie},
journal = {Opt. Express},
number = {23},
pages = {28537--28552},
publisher = {OSA},
title = {Thin sample refractive index by transmission spectroscopy},
volume = {22},
month = {Nov},
year = {2014},
url = {http://www.osapublishing.org/oe/abstract.cfm?URI=oe-22-23-28537},
doi = {10.1364/OE.22.028537},
}

@article{Poelman_2003,
	doi = {10.1088/0022-3727/36/15/316},
	url = {https://doi.org/10.1088/0022-3727/36/15/316},
	year = 2003,
	month = {jul},
	publisher = {{IOP} Publishing},
	volume = {36},
	number = {15},
	pages = {1850--1857},
	author = {Dirk Poelman and Philippe Frederic Smet},
	title = {Methods for the determination of the optical constants of thin films from single transmission measurements: a critical review},
	journal = {Journal of Physics D: Applied Physics}
}

@article{Lee:16,
author = {Choonghwan Lee and Heejoo Choi and Jonghan Jin and Myoungsik Cha},
journal = {Appl. Opt.},
number = {23},
pages = {6285--6291},
publisher = {OSA},
title = {Measurement of refractive index dispersion of a fused silica plate using {Fabry--Perot} interference},
volume = {55},
month = {Aug},
year = {2016},
url = {http://www.osapublishing.org/ao/abstract.cfm?URI=ao-55-23-6285},
doi = {10.1364/AO.55.006285},
}

@article{PETIT2020100062,
title = "Demonstration of site-selective angular-resolved absorption spectroscopy of the {$^{4}I_{15/2} \rightarrow $$^{4}I_{13/2}$} erbium transition in the monoclinic crystal {Y$_2$SiO$_5$}",
journal = "Optical Materials: X",
pages = "100062",
year = "2020",
issn = "2590-1478",
doi = "https://doi.org/10.1016/j.omx.2020.100062",
url = "http://www.sciencedirect.com/science/article/pii/S2590147820300140",
author = "Yannick Petit and Benoît Boulanger and Jérôme Debray and Thierry Chanelière"
}

@article{Traum:14,
author = {C. Traum and P. L. In\'{a}cio and C. F\'{e}lix and P. Segonds and A. Pe{\~n}a and J. Debray and B. Boulanger and Y. Petit and D. Rytz and G. Montemezzani and P. Goldner and A. Ferrier},
journal = {Opt. Mater. Express},
number = {1},
pages = {57--62},
publisher = {OSA},
title = {Direct measurement of the dielectric frame rotation of monoclinic crystals as a function of the wavelength},
volume = {4},
month = {Jan},
year = {2014},
url = {http://www.osapublishing.org/ome/abstract.cfm?URI=ome-4-1-57},
doi = {10.1364/OME.4.000057},
}

@article{li1992spectroscopic,
  title={Spectroscopic properties and fluorescence dynamics of {Er$^{3+}$} and {Yb$^{3+}$} in {Y$_2$SiO$_5$}},
  author={Li, C and Wyon, Ch and Moncorge, Richard},
  journal={IEEE journal of quantum electronics},
  volume={28},
  number={4},
  pages={1209--1221},
  year={1992},
  publisher={IEEE}
}

@article{Segonds:04,
author = {Patricia Segonds and Beno\^{i}t Boulanger and Jean-Philippe F\`{e}ve and Bertrand M\'{e}naert and Julien Zaccaro and G\'{e}rard Aka and Denis Pelenc},
journal = {J. Opt. Soc. Am. B},
number = {4},
pages = {765--769},
publisher = {Optica Publishing Group},
title = {Linear and nonlinear optical properties of the monoclinic {Ca4YO(BO3)3} crystal},
volume = {21},
month = {Apr},
year = {2004},
url = {https://opg.optica.org/josab/abstract.cfm?URI=josab-21-4-765},
doi = {10.1364/JOSAB.21.000765},
}

@article{Ghosh:97,
author = {Gorachand Ghosh},
journal = {Appl. Opt.},
number = {7},
pages = {1540--1546},
publisher = {Optica Publishing Group},
title = {Sellmeier coefficients and dispersion of thermo-optic coefficients for some optical glasses},
volume = {36},
month = {Mar},
year = {1997},
url = {https://opg.optica.org/ao/abstract.cfm?URI=ao-36-7-1540},
doi = {10.1364/AO.36.001540},
}

@article{PANG20053539_gap,
title = {Study on the growth, etch morphology and spectra of {Y2SiO5} crystal},
journal = {Materials Letters},
volume = {59},
number = {28},
pages = {3539-3542},
year = {2005},
issn = {0167-577X},
doi = {https://doi.org/10.1016/j.matlet.2005.06.036},
url = {https://www.sciencedirect.com/science/article/pii/S0167577X05006191},
author = {Huiyong Pang and Guangjun Zhao and Mingyin Jie and Jun Xu and Xiaoming He}
}

@Article{molecules30214161_FTIR,
AUTHOR = {Deka, Liza Rani and Michalska-Domańska, Marta and Mishra, Shubhra and Kshatri, D. S. and Rao, M. C. and Verma, Neeraj and Dubey, Vikas},
TITLE = {Near-Infrared Excited {Mn4+}- and {Nd3+}-Doped {Y2SiO5} Luminescent Material with Flower-like Morphology for Plant-Centric Lighting Applications},
JOURNAL = {Molecules},
VOLUME = {30},
YEAR = {2025},
NUMBER = {21},
ARTICLE-NUMBER = {4161},
URL = {https://www.mdpi.com/1420-3049/30/21/4161},
PubMedID = {41226125},
ISSN = {1420-3049},
DOI = {10.3390/molecules30214161}
}

@article{SHI2023100017,
title = {Evaluating refractive index and birefringence of nonlinear optical crystals: Classical methods and new developments},
journal = {Chinese Journal of Structural Chemistry},
volume = {42},
number = {1},
pages = {100017},
year = {2023},
note = {Special Issue for Nonlinear Optical Crystals},
issn = {0254-5861},
doi = {https://doi.org/10.1016/j.cjsc.2023.100017},
url = {https://www.sciencedirect.com/science/article/pii/S0254586123000077},
author = {Qi Shi and Lingyun Dong and Ying Wang}
}

@article{Thin_Films,
author = {Goldner, Philippe and Tallaire, Alexandre and Serrano, Diana and Tiranov, Alexey and Zhong, Tian},
title = {Rare-Earth Doped Thin Films for Optical Quantum Technologies},
journal = {Advanced Quantum Technologies},
volume = {8},
number = {11},
pages = {e2500026},
doi = {https://doi.org/10.1002/qute.202500026},
url = {https://advanced.onlinelibrary.wiley.com/doi/abs/10.1002/qute.202500026},
eprint = {https://advanced.onlinelibrary.wiley.com/doi/pdf/10.1002/qute.202500026},
year = {2025}
}

@article{AROSA2022110225,
title = {Accuracy of refractive index spectroscopy by broadband interferometry},
journal = {Measurement},
volume = {187},
pages = {110225},
year = {2022},
issn = {0263-2241},
doi = {https://doi.org/10.1016/j.measurement.2021.110225},
url = {https://www.sciencedirect.com/science/article/pii/S0263224121011349},
author = {Yago Arosa and Carlos {Damián Rodríguez-Fernández} and Alejandro Doval and Elena {López Lago} and Raúl {de la Fuente}}
}

@article{Zhong2015,
  title = {Optically addressable nuclear spins in a solid with a six-hour coherence time},
  volume = {517},
  ISSN = {1476-4687},
  url = {http://dx.doi.org/10.1038/nature14025},
  DOI = {10.1038/nature14025},
  number = {7533},
  journal = {Nature},
  publisher = {Springer Science and Business Media LLC},
  author = {Zhong,  Manjin and Hedges,  Morgan P. and Ahlefeldt,  Rose L. and Bartholomew,  John G. and Beavan,  Sarah E. and Wittig,  Sven M. and Longdell,  Jevon J. and Sellars,  Matthew J.},
  year = {2015},
  month = Jan,
  pages = {177–180}
}

@article{Rani2017,
  title = {Coherence time of over a second in a telecom-compatible quantum memory storage material},
  volume = {14},
  ISSN = {1745-2481},
  url = {http://dx.doi.org/10.1038/nphys4254},
  DOI = {10.1038/nphys4254},
  number = {1},
  journal = {Nature Physics},
  publisher = {Springer Science and Business Media LLC},
  author = {Rančić,  Miloš and Hedges,  Morgan P. and Ahlefeldt,  Rose L. and Sellars,  Matthew J.},
  year = {2017},
  month = Sept,
  pages = {50–54}
}

@article{Nicolle:21,
author = {M. Nicolle and J. N. Becker and C. Weinzetl and I. A. Walmsley and P. M. Ledingham},
journal = {Opt. Lett.},
number = {12},
pages = {2948--2951},
publisher = {Optica Publishing Group},
title = {Gigahertz-bandwidth optical memory in {Pr3$+$:Y2SiO5}},
volume = {46},
month = {Jun},
year = {2021},
url = {https://opg.optica.org/ol/abstract.cfm?URI=ol-46-12-2948},
doi = {10.1364/OL.423642},
}

@article{PhysRevLett.127.210502,
  title = {Entanglement between a Telecom Photon and an On-Demand Multimode Solid-State Quantum Memory},
  author = {Rakonjac, Jelena V. and Lago-Rivera, Dario and Seri, Alessandro and Mazzera, Margherita and Grandi, Samuele and de Riedmatten, Hugues},
  journal = {Phys. Rev. Lett.},
  volume = {127},
  issue = {21},
  pages = {210502},
  numpages = {7},
  year = {2021},
  month = {Nov},
  publisher = {American Physical Society},
  doi = {10.1103/PhysRevLett.127.210502},
  url = {https://link.aps.org/doi/10.1103/PhysRevLett.127.210502}
}

@article{Sanchez_Mejia_2026,
doi = {10.1088/2058-9565/ae1bd0},
url = {https://doi.org/10.1088/2058-9565/ae1bd0},
year = {2025},
month = {nov},
publisher = {IOP Publishing},
volume = {11},
number = {1},
pages = {015004},
author = {Sanchez Mejia, T and Nicolas, L and Gelmini Rodriguez, A and Goldner, P and Afzelius, M},
title = {Broadband and long-duration optical memory in {171Yb3+:Y2SiO5}},
journal = {Quantum Science and Technology}
}

@article{nicolas2023coherent,
  title={Coherent optical-microwave interface for manipulation of low-field electronic clock transitions in {171Yb3+: Y2SiO5}},
  author={Nicolas, Louis and Businger, Moritz and Sanchez Mejia, T and Tiranov, Alexey and Chaneli{\`e}re, Thierry and Lafitte-Houssat, Elo{\"\i}se and Ferrier, Alban and Goldner, Philippe and Afzelius, Mikael},
  journal={npj Quantum Information},
  volume={9},
  number={1},
  pages={21},
  year={2023},
  publisher={Nature Publishing Group UK London}
}

@article{Gonzalvo,
  title = {Cavity-enhanced Raman heterodyne spectroscopy in {${\mathrm{Er}}^{3+}:{\mathrm{Y}}_{2}{\mathrm{SiO}}_{5}$} for microwave to optical signal conversion},
  author = {Fernandez-Gonzalvo, Xavier and Horvath, Sebastian P. and Chen, Yu-Hui and Longdell, Jevon J.},
  journal = {Phys. Rev. A},
  volume = {100},
  issue = {3},
  pages = {033807},
  numpages = {9},
  year = {2019},
  month = {Sep},
  publisher = {American Physical Society},
  doi = {10.1103/PhysRevA.100.033807},
  url = {https://link.aps.org/doi/10.1103/PhysRevA.100.033807}
}

@article{Soharab2019,
  title = {Effect of Nd doping on the refractive index and thermo-optic coefficient of {GdVO4} single crystals},
  volume = {125},
  ISSN = {1432-0649},
  url = {http://dx.doi.org/10.1007/s00340-019-7194-z},
  DOI = {10.1007/s00340-019-7194-z},
  number = {5},
  journal = {Applied Physics B},
  publisher = {Springer Science and Business Media LLC},
  author = {Soharab,  M. and Bhaumik,  Indranil and Bhatt,  R. and Saxena,  A. and Karnal,  A. K.},
  year = {2019},
  month = Apr 
}

@inproceedings{Zelmon2005,
  title = {Optical properties of {Nd}-doped ceramic yttrium aluminum garnet},
  volume = {5647},
  ISSN = {0277-786X},
  url = {http://dx.doi.org/10.1117/12.584641},
  DOI = {10.1117/12.584641},
  booktitle = {Laser-Induced Damage in Optical Materials: 2004},
  publisher = {SPIE},
  author = {Zelmon,  David E. and Schepler,  Kenneth L. and Guha,  Shekhar and Rush,  Daniel J. and Hegde,  Shrikrishna M. and Gonzalez,  Leonel P. and Lee,  Julie},
  editor = {Exarhos,  Gregory J. and Guenther,  Arthur H. and Kaiser,  Norbert and Lewis,  Keith L. and Soileau,  M. J. and Stolz,  Christopher J.},
  year = {2005},
  month = Feb,
  pages = {255}
}

\newpage
\appendix

\section{Polarisation rotation patterns along \bax\, and \Dd.}\label{appendix:patterns}
\begin{figure}[htbp]
\centering
\includegraphics[width=.92\columnwidth]{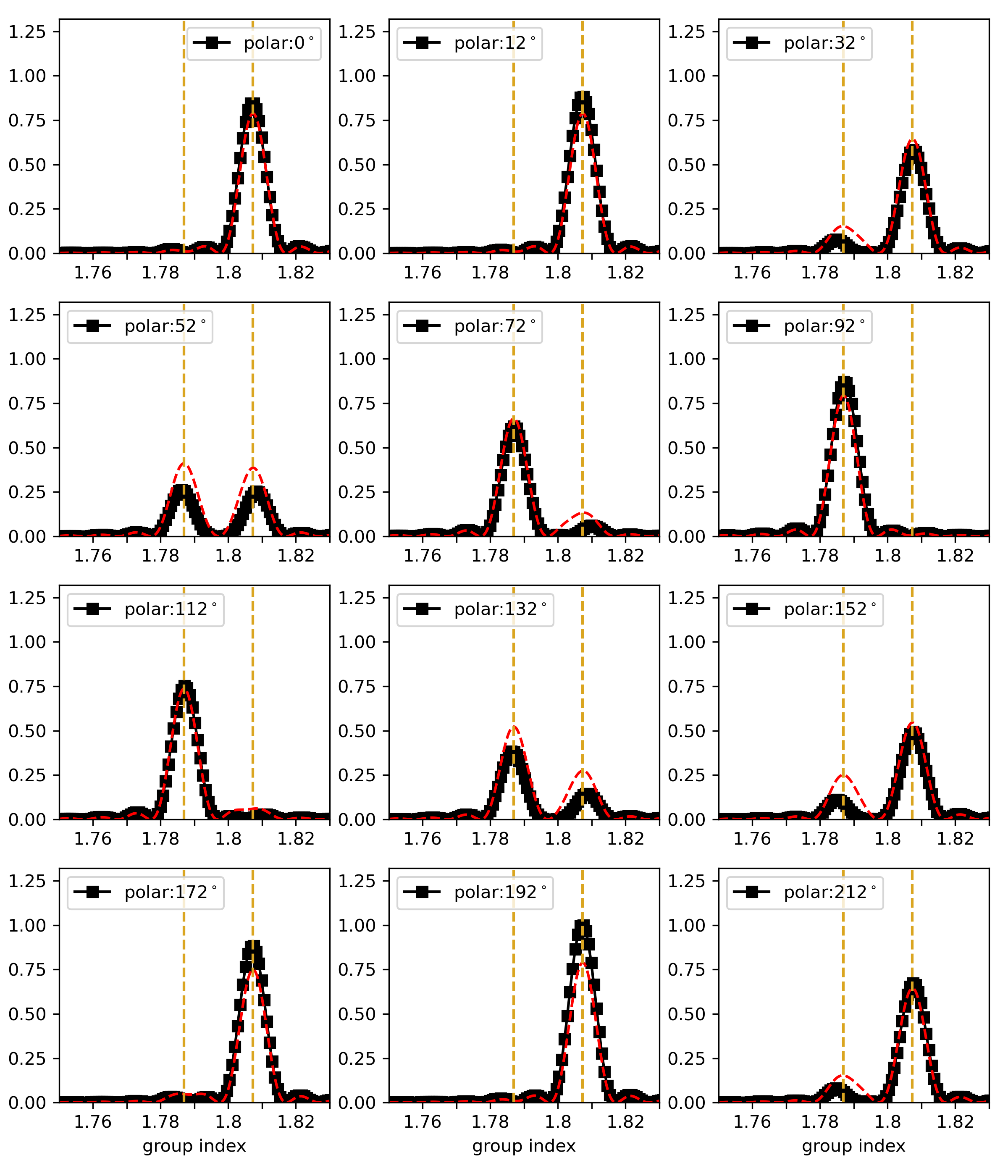} 
\caption{Absolute value of $\tilde{T}(\mathcal{N})$ (Eq. \ref{eq:Fourier} in arbitrary units) when propagating along \bax, instead of \Du\,in Fig.\ref{fig:Fourier_spectra_k_5mm}.}
\label{fig:Fourier_spectra_k_3mm}
%/home/thierry/neel_ownCloud/neel_exchange/YSO_index/20211004_YSOindex/trait_OSA_fit_global_sinc.py
\end{figure}

\begin{figure}[htbp]
\centering
\includegraphics[width=.92\columnwidth]{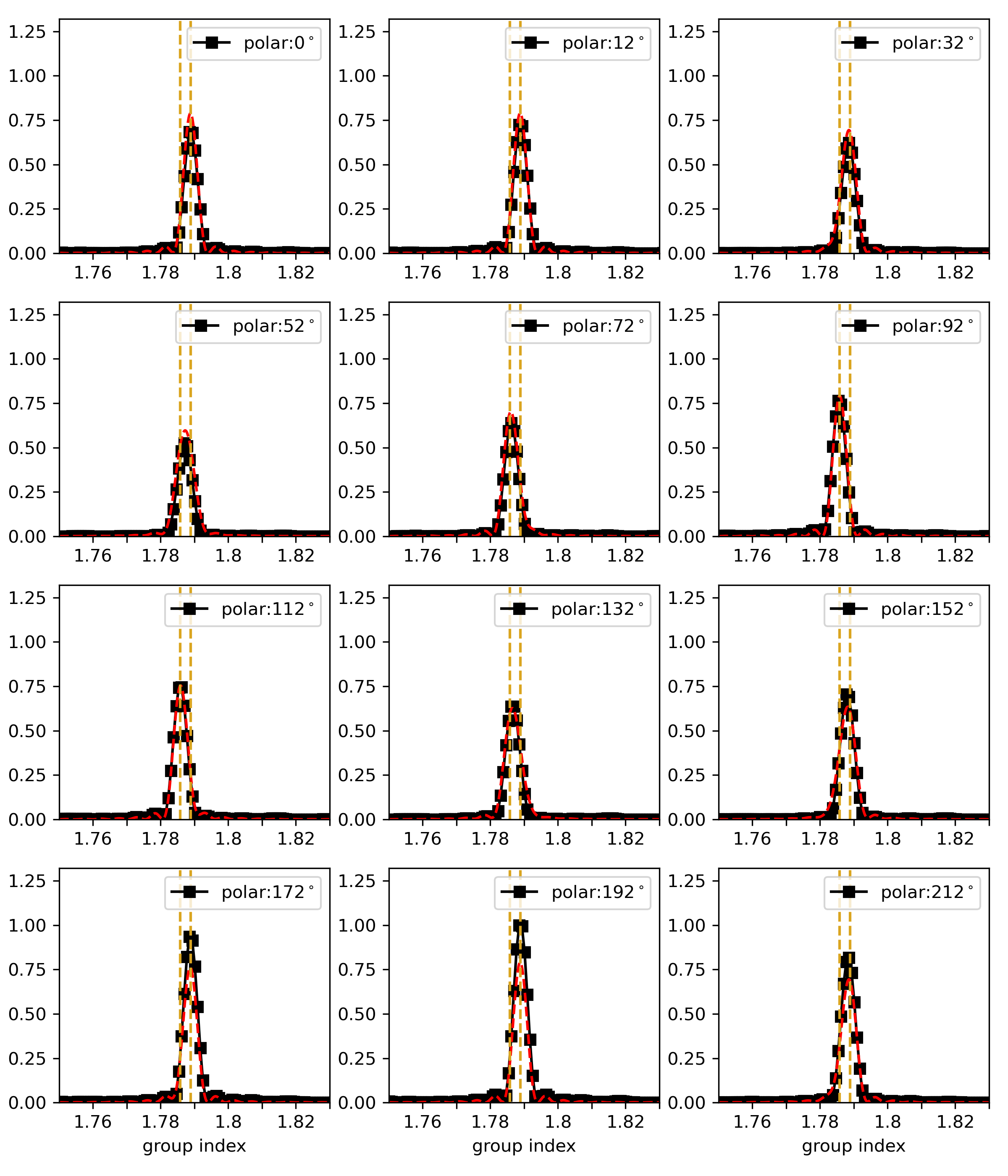} 
\caption{Absolute value of $\tilde{T}(\mathcal{N})$ (Eq. \ref{eq:Fourier} in arbitrary units) when propagating along \Dd, instead of \Du\,in Fig.\ref{fig:Fourier_spectra_k_5mm}.}
\label{fig:Fourier_spectra_k_6mm}
%/home/thierry/neel_ownCloud/neel_exchange/YSO_index/20211004_YSOindex/trait_OSA_fit_global_sinc.py
\end{figure}

\newpage
\section{Refinement of extended Sellmeier coefficients including a mid-infrared pole term}\label{appendix:Dnegative}

In Table \ref{tab:Sellmeier_new} , we proposed a set of coefficients by setting $D_i =0$. The accuracy obtained is 3.2$\times 10^{-5}$ (RMSD error). We will now repeat the fitting process, assuming that $D_i$ is non-zero, but more importantly, imposing $D_i < 0$. We obtain the following result in table \ref{tab:Sellmeier_new_Dnegative}. The precision is now 2.3$\times 10^{-5}$.

%----optim with D non-zero with bounds and Neel meas.-----
%[ 3.1085  3.1170  3.1883  0.0282  0.0285  0.0302 -0.0107 -0.0119 -0.0130
% -0.0019 -0.0002 -0.0024]
%RMSD final Neel:2.3e-05

\begin{table}[htbp]
\begin{center}
\begin{tabular}{| c |c |c |c |c| }
\hline
axis $i$ &$ A_i$ & $B_i$ & $C_i$ & $D_i$ \\
 \hline
$n_b$ &3.1085 & 0.0282  & -0.0107  & -0.0019 \\
\hline
$n_{D_1}$ & 3.1170 &  0.0285  & -0.0119 & -0.0002 \\
\hline
$n_{D_2}$ &  3.1883 & 0.0302 & -0.0130& -0.0024 \\
\hline
\end{tabular}
\end{center}
\caption{Extended Sellmeier coefficients in the near-infrared with $D_i < 0$}
\label{tab:Sellmeier_new_Dnegative}
\end{table}

The overall result makes sense, given that $C_i$ and $D_i$ are all imposed to be negative. However, the gain in accuracy is not really significant and table \ref{tab:Sellmeier_new} can be consider as sufficient. That is why we have included this refinement as an appendix. 

\end{document}